\documentclass[twocolumn,prb,floatfix,showpacs]{revtex4}
\usepackage{lscape}
\usepackage{epsf}
\usepackage[dvips]{graphicx}
\usepackage{rotating}
\usepackage{bm}
\usepackage{ulem}
\usepackage{color}
\date{\today}

\begin{document}

\title{Magnetic properties of 2D nickel nanostrips: structure dependent magnetic anomaly}

\author{Vikas Kashid}
\affiliation{Department of Physics, Savitribai Phule Pune University, Pune 411 007, India}
\author{Vaishali Shah}
\email[Email:]{vaishali@unipune.ac.in}
\affiliation{Interdisciplinary School of Scientific Computing, Savitribai Phule Pune University, Pune 411 007, India}
\author{H. G. Salunke}
\affiliation{Technical Physics Division, Bhabha Atomic Research Center, Mumbai 400 085, India}
\author{Yuriy Mokrousov}
\affiliation{Peter Gr\"{u}nberg Institut and Institute for Advanced Simulation, Forschungszentrum, J\"{u}lich and JARA, D-52425 J\"{u}lich, Germany}
\author{Stefan Bl\"{u}gel}
\affiliation{Peter Gr\"{u}nberg Institut and Institute for Advanced Simulation, Forschungszentrum, J\"{u}lich and JARA, D-52425 J\"{u}lich, Germany}


\begin{abstract}
We have investigated different geometries of two dimensional (2D) infinite length Ni nanowires of increasing width using spin 
density functional theory calculations. Our simulations demonstrate that the parallelogram motif is the most stable and structures 
that incorporate the parallelogram motif are more stable as compared to rectangular structures. The wires are conducting and
the conductance channels increase with increasing width. The wires have a non-linear behavior in the ballistic anistropic magnetoresistance ratios with
respect to the magnetization directions.  All 2D nanowires as well as Ni (111) and Ni (100) monolayer investigated are 
ferromagnetic under the Stoner criterion and exhibit enhanced magnetic moments as compared to bulk Ni and the respective Ni monolayers. The Stoner parameter is seen to depend on the structure and the 
dimension of the Nws. 
The easy axis for all nickel nanowires under investigation is observed to be along the wire axis.
The double rectangular nanowire exhibits a magnetic anomaly with a smaller magnetic moment when 
compared to Ni (100) monolayer and is the only structure with an easy axis perpendicular to the wire axis. 
\end{abstract}

\maketitle
\section{Introduction}
Magnetism in nanoscale films is an exciting area of research, as their magnetic behavior 
is different than that observed in bulk.  The applications of magnetic materials are mostly determined by the softness or hardness of the magnet. The soft magnets with low coercivities are used for flux guidance in permanent magnets and in transformer cores for high frequency and microwave applications.\cite{tung,ralph} In hard magnets, knowledge of the easy direction of magnetization and magnetic anisotropy energy is necessary for 
high density recording devices such as memories and magnetic tapes in audio-visual technology. \cite{tung,ralph} The magnetic properties of materials in the nanoscale regime are quite interesting as they show enhanced magnetic moments due to the reduced coordination, and exhibit magnetic anisotropy due to 
symmetry breaking.  

Nickel is one of the most investigated materials, for its magnetic properties and its applicability in material science. The magnetic properties of bulk nickel,\cite{Doll,Smogunov,krutzen,brookes,borgiel,Zarechnaya} Ni
 surfaces\cite{Wimmer,freeman1,freeman2,bennet,takayama,Krakauer,jepsen} and layers\cite{zhu,tersoff,Yi,pick,freeman2,gayj} have been investigated in great detail. 
Single atom linear chains and zigzag nanowires (Nws) of Ni have been studied earlier for their stability.\cite{zenley,tung,Sabirianov,Smogunov,Smogunov3,alex,Ooka,wernsdorfer,ataca} 
The investigations by Tung \textit{et al.}, on the magnetic properties of all 3$d$ transition elements have found that in Ni the linear and zigzag wires are ferromagnetic.
In a recent work, Zelen\'{y}  \textit{et al.},\cite{zenley} have investigated systematically the variation in structural and magnetic properties of one dimensional nanowires, two dimensional strips and three dimensional rods, under compression. The compressed two dimensional (2D) Ni nanowires have a ferromagnetic ground state, because of magnetic shape anisotropy\cite{tung,zenley}. This result contradicts the work of Mermin-Wagner\cite{mermin}, which showed that, one and two dimensional structures cannot exhibit ferromagnetic or antiferromagnetic behavior, under the isotropic Heisenberg model. Clearly, the magnetic anisotropy plays a crucial role which needs to be understood and investigated for a fundamental understanding of the magnetic behavior and its 
dependence on the structure. The magnetic anisotropy in low dimensional chains is seen to exhibit significantly different behavior than bulk.\cite{tung} The results for linear (viz., V, Cr and Fe) as well as zigzag (Cr, Mn and Co) transition metal chains show that these systems show stable axial magnetic anisotropy energy, with large magnitude in comparison with that of bulk. It was also observed that, the magnetic anisotropy energy especially, shape anisotropy energy was sensitive to the atomic arrangement of the nanowires even in low dimension. 
These results highlight the strong impact of magnetic anisotropy energy on the magnetic ground state of low dimensional material. Nanowires with very strong magnetic anisotropy with the magnetic moments in the ferromagnetic order have potential applications in ultrahigh density magnetic memories and storage devices.\cite{tung} Hence, the relationship between magnetic anisotropy energy and the geometrical structure of the atomic chains with gradually varying atomic width will be interesting for basic understanding as well as for specific applications in devices.  This kind of an investigation is not available either theoretically or experimentally for 2D nanostrips. 
In addition to the magnetic ground state, the magnetic anisotropy and hence the direction of the magnetic moments has a significant effect on the ballistic conductance in magnetic nanowires. The experimentally observed ballistic conductance in linear Ni Nws was successfully explained by considering the magnetic domain formation over uniform magnetic moments.\cite{smogunov1,alex,Ooka,sirvent,li,garcia,mochales} The exact effect of orientation of the magnetic moment with respect to the current direction on the ballistic conductance in the linear Ni Nw was demonstrated by Velev \textit{et al,}\cite{velev} and it was found that changing the direction of the magnetic moment either  perpendicular  or parallel  to the current direction enhances the ballistic conductance upto 14\%. 
Hence, it will be very interesting to investigate the effect of the direction of the magnetic moment on the ballistic conductance of the nanowires with reduced symmetry and larger widths than the linear nanowire.

In the present paper, we have investigated the magnetic and electronic properties of nickel nanowires as their structure varies from a linear chain to two dimensional nanostrips of increasing widths. 
Besides their stability, we have investigated the effect of the geometrical structure on their magnetic properties. We confirm the ferromagnetic ground state of Ni Nw structures based on Stoner's criterion. We have studied systematically the the effect of geometrical structure on the spin dependent ballistic conductance. We observe a structure dependence on the value of the Stoner's integral which is so far known to be a constant for a particular material. The structure of the paper is as follows: Section II gives an overview of computational parameters and methods used. The results on the stability, electronic structure, conductance, magnetic properties and magnetic transition are discussed under Sec. III which is followed by a summary of the 
work in Sec. IV.

\section{Computational Details}

\begin{figure*}

\begin{center}
{\scalebox{0.37}{\includegraphics*{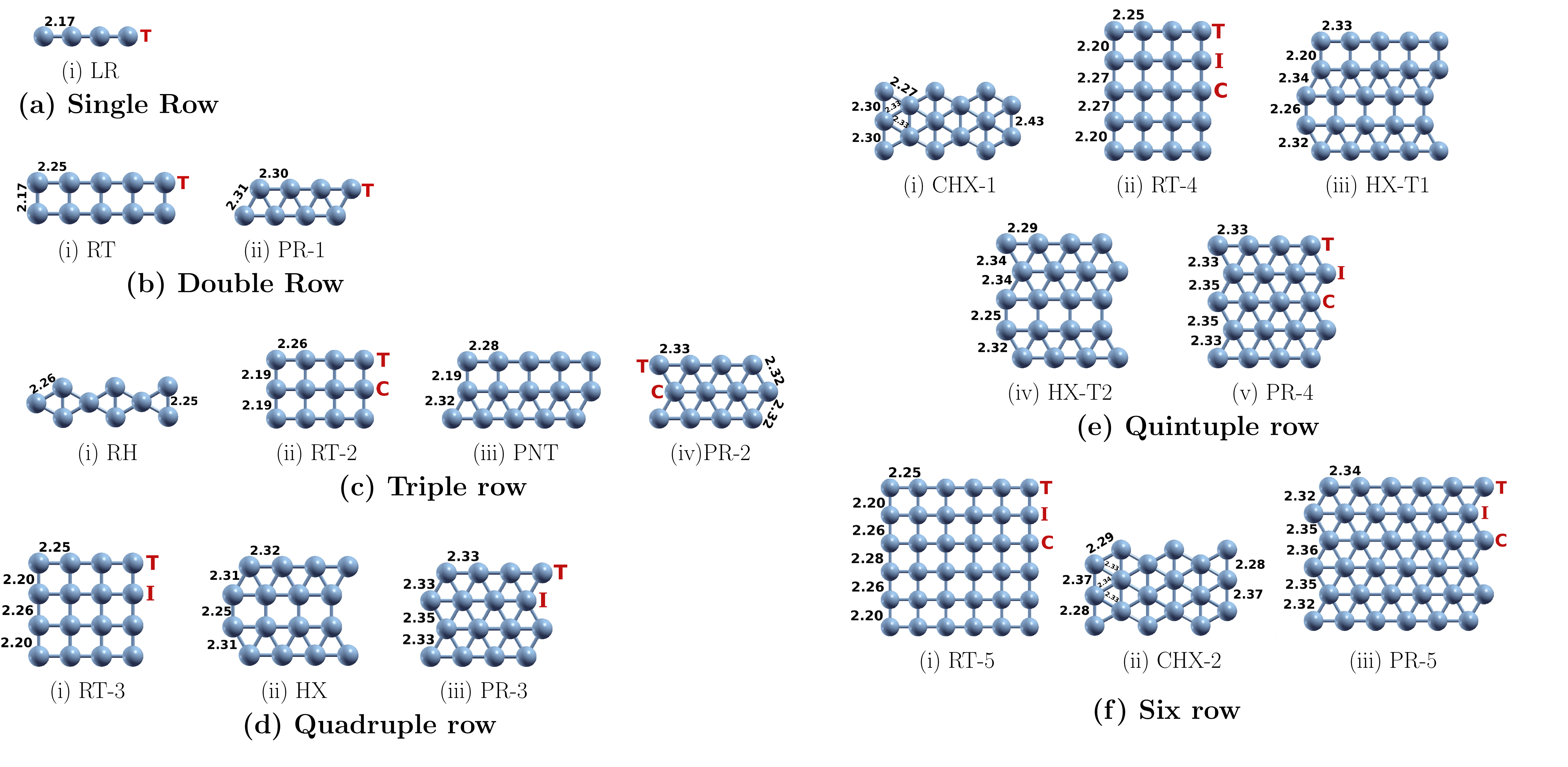}}}\\
\end{center}
 \caption{The optimized geometries of the nickel nanowires with increasing number of rows (1-6) of atoms. In each row, the geometries are arranged in ascending order of stability. `T', `I' and `C' denote the atoms representing the terminal, intermediate and the central row of the nanowire from the vacuum end to the center of the nanowire along the non-periodic direction.}
    \label{f-geom}
\end{figure*}
We have performed total energy calculations based on \textit{ab initio} spin density functional theory (DFT) 
as implemented in the  Vienna \textit{ab initio} Simulation Package(VASP) \cite{vasp1, vasp2, vasp3}. The  Projector Augmented Wave method (PAW) \cite{bloch} and the Generalized Gradient Approximation (GGA)\cite{perdew} have been used to describe the electron-ion interactions and the exchange correlation interactions.   
The details of modeling of such nanowires and the calculation of the binding energy have been described earlier for gold.\cite{vikas}
The geometry optimization and the electronic band structure calculations are performed without the scalar relativistic effects, as the investigations on 3$d$ transition metals have demonstrated that
the spin orbit coupling energy is approximately 3 times smaller than the exchange splitting in bulk Ni,\cite{Wimmer} and hence can be neglected. 

The magnetic anisotropy in a magnetic material has contributions from magnetocrystalline anisotropy and magnetic shape anisotropy. In bulk Ni, the shape anisotropy energy does not contribute and the only contribution is from the magnetocrystalline anisotropy. However, in 1D wires and 2D strips, the reduced symmetry of the system gives non-zero contributions to the shape (magnetostatic) anisotropy.
The magnetic shape anisotropy is computed based on the magnetic dipolar energy, resulting from the interactions of the magnetic moments with each other. It is given by
\begin{equation}  
 \varepsilon_{d} = \mu_{B}^{2} \sum_{i,j}{}^{'} \left( \frac{{\mathbf{m}_{i}}\cdot \mathbf{m}_{j}}{ \vert  \mathbf{r}_{ij} \vert ^{3} }  - \frac {3 ( \mathbf{m}_{i} \cdot \mathbf{r}_{ij}) (\mathbf{m}_{j}\cdot \mathbf{r}_{ij})} {\vert \mathbf{r}_{ij} \vert ^{5}} \right)  
\label{eq-mag-dip}
\end{equation}
\noindent
where, \textbf{m}$_{i}$ and \textbf{m}$_{j}$ are the magnetic moments at the ionic positions \textbf{r}$_{i}$ and \textbf{r}$_{j}$. The vector \textbf{r}$_{ij}$ is defined as $\vert$\textbf{r}$_{i}$-\textbf{r}$_{j}\vert$.
The magnetic dipolar energy $\varepsilon_{d}$ is a long range interaction, hence the value of $\varepsilon_{d}$ has been converged over the length of the nanowire by using supercell. The dipolar energy is sensitive to the orientation of the magnetic moments, hence we define magnetic shape anisotropy (MSA) as, $K^{s}_{1}=\varepsilon_{d}^{\hat{y}}-\varepsilon_{d}^{\hat{x}}$ and  $K^{s}_{2}=\varepsilon_{d}^{\hat{z}}-\varepsilon_{d}^{\hat{x}}$ where $\varepsilon_{d}^{\hat{x}}$, $\varepsilon_{d}^{\hat{y}}$ and $\varepsilon_{d}^{\hat{z}}$ are the classical dipole-dipole interaction energy terms, when the magnetic moments are aligned along $\hat{x}$, $\hat{y}$ and $\hat{z}$ direction.
 
For the calculation of the magneto-crystalline anisotropy energy (MCA), we have done self consistent calculations by including the spin orbit coupling and with the magnetization direction oriented along one of the Cartesian axes. We have used the total energy difference method for the calculation of the anisotropy energy (MCA) due to the orientation of the magnetization along the different directions. 
We have defined
$K^{c}_{1}$= E$^{\hat{y}}$-E$^{\hat{x}}$, and $K^{c}_{2}$=E$^{\hat{y}}$-E$^{\hat{z}}$, where, $\hat{x}$ is the periodic wire direction, $\hat{y}$ is the perpendicular direction in the plane of the wire and $\hat{z}$ is perpendicular to both the wire axis and the wire plane. The total magnetic anisotropy (MAE) is calculated as, $K_{1(2)}=K^{s}_{1(2)}+K^{c}_{1(2)}$.
 
In order to understand the electronic and conduction properties of the nanowires, we have performed band structure analysis as described by Kashid \textit{et al.} \cite{vikas}
The conducting properties of nickel nanowires are investigated based on
the Landuaer-B\"{u}ttiker formalism, \cite{landauer} using the method of plane waves. \cite{oetzel,oetzel2}
For ideal infinite length nanostrips, the ballistic conductance ($G$) can be written as,  
\begin{equation}
G=\frac{2 e^{2}}{h} N(E) 
\label{eq-cond-1}
\end{equation}
where, $(2 e^{2}/h)$ represents the quantization unit of conductance, in the case of spin degeneracy. The term $N(E)$ shows the number of conductance channels, and it is explicitly calculated by counting the number of bands crossing the Fermi energy E$_{F}$). For ferromagnetic nanowire, the conductance is expected to exhibit spin dependence, where spin up ($\uparrow$) and spin down ($\downarrow$) electrons contribute independently to the number of conductance channels.\cite{Ooka} Hence Eq. \ref{eq-cond-1} can be modified for ferromagnetic nanowires as,
\begin{equation}
G=\frac{e^{2}}{h} \Big(N_{\uparrow}(E) + N_{\downarrow}(E)\Big)
\label{eq-cond-2}
\end{equation}


where, $N_{\uparrow (\downarrow)}(E)$ is the number of spin up (down) bands at the Fermi energy. 
The effect of spin orbit coupling (SOC) on the conduction in Ni Nws has been investigated based on the ballistic anisotropic magnetoresistance ($BAMR$) computation described by Velev \textit{et al.}\cite{velev} We have calculated the percentage $BAMR$ as, \\
\begin{equation}
 BAMR = \frac{N_{\perp} - N_{\parallel}}{N_{\perp}} \times 100
\label{eq-bamr}
\end{equation}

where $N_{\perp}$ is the total number of conducting channels, when the magnetization is perpendicular to the current direction (axial direction) and $N_{\parallel}$ is the number of conducting channels, when the magnetization is parallel to the axial direction. 
The two anisotropic magnetization directions viz., $\hat{y}$ and $\hat{z}$ that are perpendicular to the current direction $(\hat{x})$ give different percentage BAMR denoted by A (B), respectively.\\

We have used the Stoner's model of ferromagnetism to determine the appearance of ferromagnetic ground state in Ni Nws. The product satisfying $I \cdot n^{\circ}_{E_{F}} > 1$ defines the system as  ferromagnetic, where $I$ is the Stoner exchange parameter, obtained from $\Delta_{ex}=I m$. The $\Delta_{ex}$ defines the exchange splitting between majority and minority spin electrons due to magnetization, and is calculated directly from the total density of states (DOS) of nickel nanowires. The quantity $m$ denotes the magnetic moment per atom, whereas $n^{\circ}_{E_{F}}$ denotes the normalized DOS (DOS per atom per spin) at the Fermi energy ($E_{F}$) of the nonmagnetic calculation.

\section{Results and Discussion}

\subsection{Structure and stability}

We have investigated different geometries of Ni nanostrips of widths upto 1 nm approximately. Figure \ref{f-geom} shows the  optimized geometries of nickel nanowires arranged according to increasing number of rows. In each row, the structures are arranged in the ascending order of stability. The bond lengths between nickel atoms in the unit cell of each nanowire are displayed in Fig. \ref{f-geom}.  
Linear Ni nanowire (Fig. \ref{f-geom}a(i)) has been investigated earlier in detail\cite{Zarechnaya,tung,smogunov1,Ooka,Wen,li,Smogunov2} and we report it for comparison with the magnetic
properties of increasing width nanostrips. Our bond length of 2.17 \AA~ is in good  agreement with the calculations of Zelen\'y \textit{et al.} \cite{zenley}
Like gold Nws, infinite length linear Ni nanowire does not exhibit dimerization in bond lengths.\cite{vikas}

\begingroup
\begin{table*}
\caption{The width, bonding factor, magnetization and spin resolved conductance channels for nickel nanowire geometries shown in Fig. \ref{f-geom}. The nanowires are arranged according to increasing number of rows. In each row, structures are listed in increasing order of stability. }
\begin{tabular}{l c c c c c c c } \hline \hline
Geometry & Number \hskip 0.18in & Width & Bonding Factor & Magnetization & Conduction & channels  \\ 
     & of rows & (nm)   & (bonds/atom) & $(\mu_{B}$/atom) &  ($\uparrow$) &  ($\downarrow$)\\ 
\hline

Linear (LR) & $1$ & $0.00$ &1.00 & $1.16$ & 1 & 6 \\

Rectangular (RT-1)& $2$&  $0.22 $  &1.50 & $0.90$ & 2 & 6\\

Parallelogram (PR-1)& $2$& $0.20$  & 2.00 & $0.94 $ & 2 & 6 \\

Rhombus (RH) & $3$ &  $0.23$   & 1.33 & $ 0.75$ & 1 & 3 \\

Double rectangular (RT-2) & $3$ & $0.44$  & 1.66 & $0.81 $ & 3 & 7 \\

Pentagonal (PNT) & $3$ &  $0.42$ & 2.00 & $0.82 $ & 2 & 6 \\ 

Double parallelogram (PR-2) & $3$ &  $0.40$ & 2.33 & $0.86$ & 2 & 5 \\ 

Triple rectangular (RT-3) & $4$& $0.67$  &1.75 & $0.83 $ & 4& 9  \\

Hexagonal (HX) & $4 $ &$0.62$    &2.25 & $ 0.88$ &3 & 7 \\

Triple parallelogram (PR-3) & $4$ &  $0.61$ & 2.50 & $0.86 $ & 3 & 10 \\

Centered hexagonal (CHX-1) & $5$& $0.46$  & 2.20 & $ 0.86$ & 2 & 6\\

Quadruple rectangular  (RT-4) & $5$& $0.89$  & 1.83& $0.86 $ & 5 & 12 \\

Hexagonal type-I (HX-T1) & $5 $ & $0.88$  & 2.20 & $0.84 $  & 4 & 12 \\

Hexagonal type-II (HX-T2) & $5 $ & $0.83$  & 2.39& $0.86 $ & 4 & 12\\

Quadruple parallelogram (PR-4) & $5$ &  $0.81$ & 2.60 & $0.85 $ &4 & 13 \\ 

Quintuple rectangular (RT-5) & $6 $ & $1.12$  &1.80 & $0.89 $ &6 & 17 \\

Centered hexagonal-II (CHX-2)&$6$ &$0.58$ & 2.16 & $0.82 $ & 3 & 4 \\

Quintuple parallelogram  (PR-5) & $6$  & $1.02$  & 2.66 & $ 0.85 $ & 4 & 12 \\

Bulk nickel & $ $  &   & $6.00$ & $ 0.62 $ &  &  \\
\hline
\end{tabular}
\vskip 0.1in
\label{t-bind}
\end{table*}
\endgroup

In the double row Nws, nickel forms either rectangular or parallelogram structure.  
Our investigations show that, unlike gold nanowires (GNWs), Ni Nws show a single  minima in the double row parallelogram structure, which consists of all acute angle triangles.
 The double row parallelogram and rectangular structure form the basic building blocks for all subsequent higher row structures.
The important difference between GNWs and Ni Nws is that, the metastable structures like zigzag (ZZ), hexagonal (HX), double zigzag (DZZ) in GNWs are not observed in Ni Nws.
The optimization of these structures leads to more stable PR-1 and HX structure (that is formed of a rectangular structure sandwiched between two parallelogram structures).
In the present work, we have investigated additional structures like HX-T1 and HX-T2 (Fig. \ref{f-geom}e(iii) and Fig. \ref{f-geom}e(iv) respectively), which are formed by different combinations of rectangular and parallelogram structures.
 HX-T1 consists of an alternate parallelogram and rectangular row of Ni atoms along the Y direction, whereas, in the HX-T2 case, the topmost rectangular row is replaced by a parallelogram row. 
These additional structures assist to understand that, the stability in Ni NWs is strongly dependent on the geometrical arrangement of Ni atoms. 
All rectangular nanowires show a bond length of 2.25 \AA~ along the periodic direction except for RT-2 structure which has a 2.26 \AA~ bond length. 
The parallelogram nanowires have slightly enhanced bond lengths of 2.30 \AA~ for PR-1, PR-2,  2.33 \AA~  for PR-3, PR-4, and  2.34 \AA~  for PR-5 structure, in comparison with that of rectangular nanowires. These bond lengths are converging to 2.26 \AA~ and 2.36 \AA, respectively$-$the single layered limit for Ni (100) and Ni (111) films.
The rectangular and parallelogram nanowires show dimerization in bond lengths along the nonperiodic direction, similar to that observed in GNWs.\cite{vikas}
 The dimerization in bondlengths is observed for nanowire structures made up of 4 or more rows of nickel 
 atoms and it is symmetric from the center of the structure along the nonperiodic Y direction.\\

\begin{figure}
    \includegraphics[angle=270,width=3.10in]{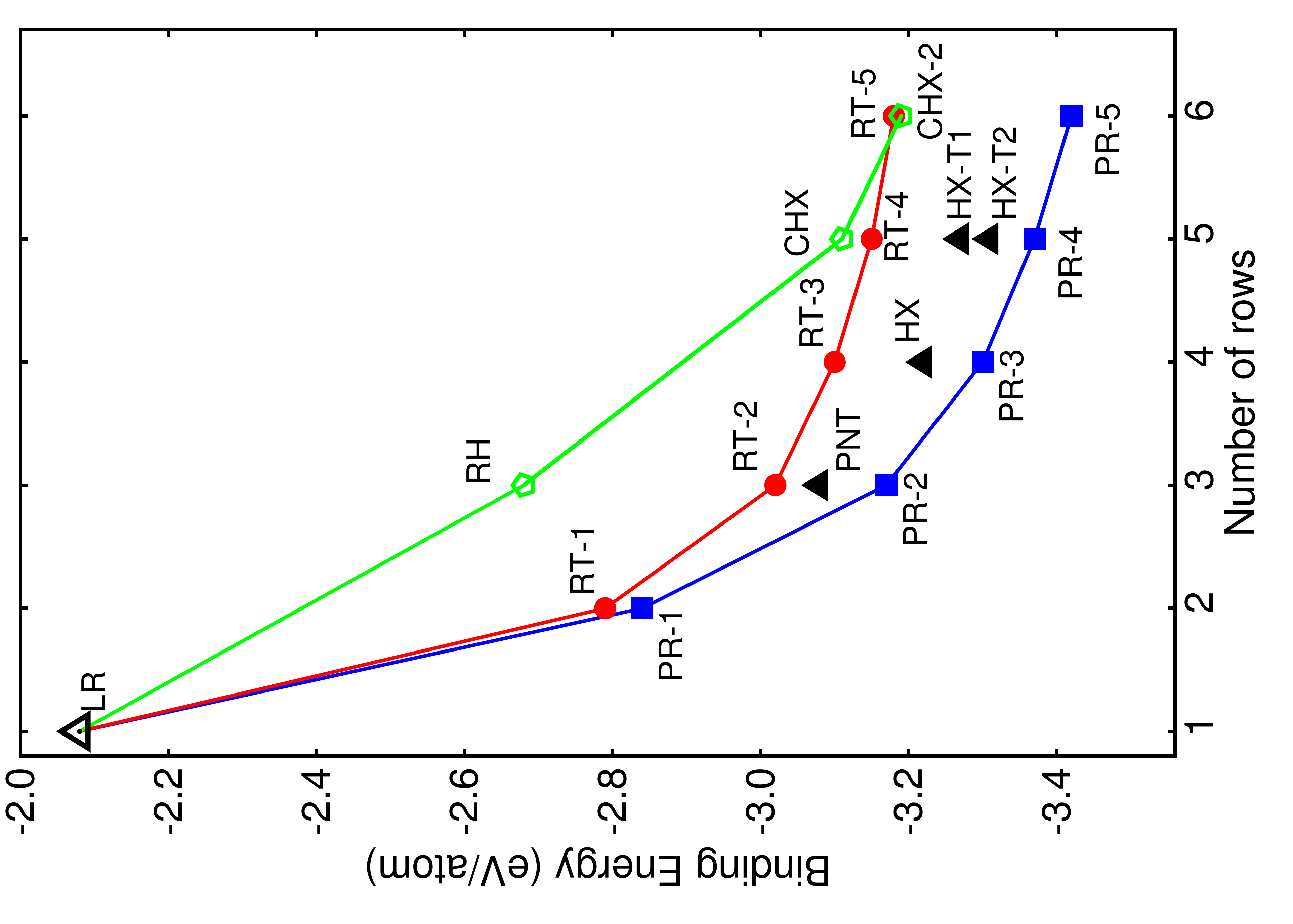}
    \caption{The binding energy as a function of the number of rows of atoms in the Nw structure.
The filled squares represent parallelogram structures, filled circles represent rectangular structures. The open pentagons represents rhombus structure, and all other remaining structures are shown by triangles.}
    \label{f-bind}
\end{figure}

Figure \ref{f-bind} shows the plot of the calculated binding energies per atom as a function of number of rows,
 for each nanowire investigated. In general, as the width of the nanostrips 
increases the stability of the structures increases. The parallelogram shaped nanowires are the most
stable structures within a particular number of rows. Nanowires formed via a combination of parallelogram 
and rectangular motifs are stable over pure rectangular shaped structures. The energy difference between rectangular shaped and parallelogram shaped structures, is small at smaller widths, however, the difference increases with an increase in the width. Our calculations show that a single layer parallelogram film is 
more stable than rectangular film by 0.34 eV/atom. This increase in the energy difference between rectangular and parallelogram nanowires with increasing number of rows was observed previously in the case of GNWs.\cite{vikas}
The energy difference between rhombus and rectangular wires is large for 3 rows, however, the difference reduces with increasing number of rows and crossover in stability is observed at 6 rows.
The rhombus structure and rectangular structures are the least stable. The structures that incorporate parallelograms are energetically more preferred. The HX-T2 structure with more rows of parallelograms is 
stable over HX-T1 structure. The binding energies of the rectangular and parallelogram shaped nanowires are converging to their 
calculated single layered film limit of --3.34 eV/atom and --3.68 eV/atom, respectively.\\  

\begin{table*}
\caption{The l-decomposed charges per atom within the sphere having Wigner-Seitz radius, for the terminal (T), intermediate (I) and central (C) atoms of rectangular and parallelogram nickel nanowires shown in Fig. \ref{f-geom}.}
\begin{tabular} {l c  c  c  c  c  c  c  c  c  c} \hline \hline
Rows & \hspace {0.1in} site \hspace {0.1in} &  \multicolumn{4}{c} { Rectangular Wires } &   &   \multicolumn{4}{c} {Parallelogram Wires }  \\ \cline{3-6}  \cline{8-11}
in Nws  &       & \hspace {0.2in} $s$ \hspace {0.2in} & \hspace {0.2in}  $p$ \hspace {0.2in} & \hspace {0.2in} $d$ \hspace {0.2in}  & \hspace {0.2in} total \hspace {0.2in} & \hspace {0.3cm} & \hspace {0.2in} $s$ \hspace {0.2in} & \hspace {0.2in} $p$ \hspace {0.2in} & \hspace {0.2in} $d$ \hspace {0.2in}  & \hspace {0.2in} total \hspace {0.2in} \\ \hline
\hline

1  & T  & 0.55  &0.15  &8.33  &9.03  &   & 0.55 & 0.15 & 0.83 & 9.03 \\
\hline
2  & T  & 0.54  & 0.23 & 8.35 &9.12  &   &0.53  &0.25  &8.35  & 9.12 \\
\hline
3  & T  & 0.54  & 0.21 & 8.37 & 9.12 &   & 0.54 & 0.22 & 8.36 & 9.12 \\

  &  C & 0.54  &0.34  & 8.34 & 9.22 &   & 0.54 & 0.39 & 8.31 & 9.24 \\
\hline
4  & T  & 0.54  &0.21  &8.36  & 9.11 &   & 0.54 & 0.22 & 8.35 & 9.11 \\

  &  C & 0.54  & 0.32 & 8.35 & 9.21 &   &0.55  &0.38  &8.30  &9.23  \\
\hline
5  & T  & 0.54  & 0.21 & 8.36 & 9.11 &   & 0.54 & 0.22 &8.35  & 9.11 \\

  &  I  & 0.54  & 0.32 & 8.34 & 9.19 &   & 0.55 & 0.37 & 8.30 & 9.23 \\

  & C  & 0.53  & 0.30 &8.35  &9.18  &   & 0.56 & 0.37 & 8.31 & 9.23 \\
\hline
6  & T  & 0.54  & 0.21 & 8.35 & 9.10 &   & 0.54 &0.22  &8.35  & 9.11 \\

  &  I  & 0.54  & 0.32 & 8.34 & 9.20 &   & 0.55 &0.37  & 8.31 & 9.22 \\

  &  C  & 0.54  & 0.30 & 8.35 & 9.18 &   & 0.56 & 0.35 & 8.31 & 9.22 \\
\hline
\end{tabular}
\vskip 0.1in
\label{t-charge}
\end{table*}

In order to understand the stability of the nanowires, we have calculated the average number of nearest neighbors in each structure and listed them as bonding factor in Table-\ref{t-bind}, along with the number of rows and 
the width of the structure. 
The higher stability of parallelogram Nws in any row can be attributed to their higher bonding factor (or largest number of nearest neighbors) in comparison with the other structures that have the same number of rows. 
In addition, we have also analyzed the decomposition of total charge in the parallelogram
and rectangular structures and listed their \textit{l}-projected components in Table \ref{t-charge} along with that of the linear 
wire for comparison. 
The values of the charges are listed for the atoms along the $y$-direction from the vacuum end to the center of 
the nanowire. 
The atomic sites for the terminal row, intermediate row and central row atoms are labeled as `T', `I' and `C', as shown in Fig. \ref{f-geom}.  
It is observed that the least coordinated linear structure has the least amount of charge in $p$ orbitals as compared to that in all the 
higher width Nws.
Adding one row of Ni atoms to the linear wire structure, increases the electron occupation in $p$ orbitals of the terminal atoms to 0.08 
and 0.10 for the rectangular and parallelogram Nws, respectively.

For all the Nws with number of rows 
$\geq$ 3, the increased coordination of the `I' and `C' kind of atoms enhances the $p$ orbital occupancy. 
In the rectangular shaped nanowires with 5 and 6 rows of atoms, the central atoms loose some of their $p$ charge 
to the interstitial region, thus reducing the total charge of the `C' atoms.
Interestingly, the `I' atoms of 5 and 6 row rectangular atoms show more total charge than `T' and `C' atoms. The `C' atoms of 5 and 6 row loose more charge to the vacuum in comparison with `I' atom, although both have the same coordination, unlike that of the parallelogram Nws of the same number of rows.
In parallelogram Nws, only the terminal row atoms are affected by the metal-vacuum interface in the 
XY-plane and all sub-terminal atoms (`I' and `C' type) have essentially unchanged total charge. 
This reduction in $p$ charge of the `T' atoms at the metal vacuum interface of two dimensional films
is analogous to the loss of $p$ charge by the surface atoms observed by  Wimmer \textit{et al}. 
for a 7 layer slab of Ni (001).\cite{Wimmer}
Our results demonstrate that the metal vacuum interface has the largest effect on the $p$ orbital electron distribution.
Increase in total charge on an atom in parallelogram Nws is because of the increase in $p$ contribution. This increase in total charge as compared to that of rectangular Nws enhances the stability of parallelogram Nws. We believe this stability preference will be retained in nanowires deposited on surface.

\subsection{Electronic structure and conductance}
We have calculated the spin-dependent electronic energy band structure (BS) and density of states (DOS) of all the nanowires to
study their electronic and conducting properties.  As the number of rows in the Nws is increased, the $d$ states are increasingly 
delocalized leading to an increase in the stability of the nanowires of increasing widths. 
In all the Nws investigated here, the 3$d$, 4$s$ and 4$p$ states hybridize, so that all the spin up states are filled, 
while the spin down states are left unfilled. The spin dependent band 
structure shows large number of unfilled spin down energy bands in comparison with the spin up bands at the Fermi level. 
In the spin up band structure, the $d$ orbital bands are approximately 0.5 eV below the Fermi level. 
The fat band analysis of the spin up band structures of all nickel nanowires shows contributions at the Fermi energy from $s$, $d_{xy}$ and  $d_{x^{2}-y^{2}}$ orbitals. The contributions from the $s$ states are large in comparison with the contributions from the $d$ states. 
Amongst the $d$ states,  $d_{x^{2}-y^{2}}$ contributes strongly and $d_{xy}$ has minor contributions in the rectangular Nws. 
In parallelogram Nws, the  $d_{x^{2}-y^{2}}$  orbitals are filled, pulled below the Fermi level and $d_{xy}$ contributes strongly 
at the Fermi level. 
In rectangular nanowires as the number of rows is increased, the $d_{x^{2}-y^{2}}$ states extend beyond the Fermi energy 
leaving them increasingly unfilled. As the number of rows in parallelogram nanowires is increased, the $d_{xy}$ orbitals become more 
delocalized and their contribution to the spin up states increases. Our analysis from the partial DOS reveals that this increased 
contribution is from the atoms in the inner rows of the Nw structure. The terminal rows of atoms do not give significant contribution 
from the $d_{xy}$ orbital. 
As the coordination of the atoms in the nanowires increases with increasing widths, the $p_{x}$ and $p_{y}$ orbitals give enhanced 
contributions in spin up bands, which assists to increase the stability of the nanowires. 
On the other hand, in the spin down energy band structures, all 
the $d$ orbitals show dominant contributions along with the $s$ orbital at the Fermi level. The hybridized spin down energy levels have 
less contributions from the $p$ electrons.

The conductance in nickel nanowires was calculated using Landauer formula\cite{landauer} (Eq. \ref{eq-cond-1} and Eq. \ref{eq-cond-2}). 
Our calculated values of conductance in all nickel nanowire are listed in Table \ref{t-bind} for spin up and spin down channels. 
The less stable nanowires show higher conductance than the more stable nanowires, in agreement with the conductance behavior in Au Nws. 
All rectangular nanowires with rows $\geq$ 2 have enhanced number of conduction channels than that of parallelogram nanowires. 
The enhanced conductance in rectangular Nws in comparison with parallelogram Nws was also observed in Au nanowires.\cite{vikas}
All rectangular nanowires show a consistent increment in the total number of (spin up and spin down) conduction channels with an increase
in the width of the nanowires. However, the parallelogram Nws show a very small increase in total number of conduction channels as the
width increases. It must be noted that in all Nws, there is not a significant increment in the number of spin up channels as the number of rows is increased, however, there is a large enhancement in the number of spin down conduction channels. 
 In agreement with earlier calculations,\cite{alex} the
charge carriers in spin up conductance channels are mainly $s$ type and 
in spin down conductance channels the charge carriers are mostly $d$ type
with minor contributions from $s$ type.

\subsection{Magnetic Properties}
\subsubsection{Magnetic moment}\label{sec-magmom}
\begin{figure}
   \includegraphics[width=2.8in]{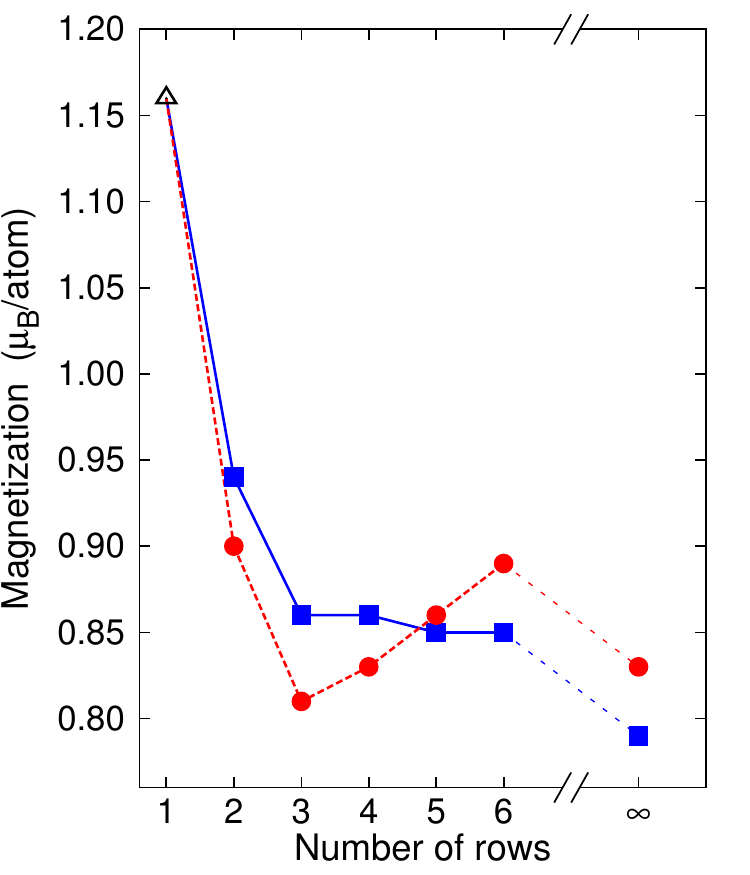}
    \caption{The magnetization in 2D nickel nanowires with increasing number of rows of atoms. Filled squares represent parallelogram nanowires, filled circles represent rectangular nanowires. Magnetization in linear Nw is shown by triangle.}
    \label{f-mag}
\end{figure}
Earlier studies on Ni surfaces have shown that (111) films of $> 3$ layers are ferromagnetic, whereas
films of $< 3$ layers show a paramagnetic behavior.\cite{bergmann,Pourovskii} This raises an interesting
question, what is the effect of the width and structure of single layered nanostrips on their 
magnetic properties? 
With this aim, we have investigated the magnetic properties of Ni Nws and 
compared them with Ni monolayers. The calculated magnetic 
moments of the Nws are listed in Table \ref{t-bind}. As expected, in nanostructures 
with reduced coordination, all Nws have enhanced magnetic moments in comparison with that of the bulk. 
The linear wire with the least coordination has the largest magnetic moment and the magnetic moments decrease with minor 
fluctuations as the widths of the nanostrips increase. A comparison of the magnetic moments of the rectangular and 
parallelogram shaped wires is plotted in Fig. \ref{f-mag}, and it shows that the magnetic behavior of these wires 
is considerably different. In the rectangular wires, the magnetic moments initially decrease and then 
increase again. Interestingly, the triple row RT-2 structure, has a magnetic 
moment of 0.81 $\mu_{B}$/atom, which is 
less than the magnetic moments of other higher row rectangular structures as 
well as that of the Ni (100) monolayer. 
Our calculated magnetization of 0.83 $\mu_{B}$/atom for a (100) monolayer film, is in good agreement with 0.86 $\mu_{B}$/atom of (100) monolayer.\cite{freeman1} 
The magnetic behavior of RT-2 is unusual when 
compared to the observation that magnetic moments increase as the atomic coordination decreases.  
We find that the ``magic structure'' RT-2 has an uncommon behavior in all the magnetic aspects 
analyzed. The parallelogram shaped wires exhibit an initial sharp decrease and then the magnetic 
moments decrease very gradually with small fluctuations.  The six row parallelogram structure (PR-5) has 
a magnetization of 0.85 $\mu_{B}$/atom, which is larger than the calculated (111) film 
value of 0.79 $\mu_{B}$/atom. Our calculated bulk magnetization value of 0.62 $\mu_{B}$/atom is in good 
agreement with the earlier reports.\cite{freeman1,zhu,dal,tersoff}

The values of decomposed magnetic moments for all symmetric atoms (described as `T', `I' and `C' in the 
previous subsection) in rectangular and parallelogram Nws are listed in Table-\ref{t-spin}. 
As expected for magnetic systems, the difference in the contributions from spin up $d$ orbitals and the spin down $d$ 
orbitals, gives rise to the spin imbalance in Nws. 
However, the spin imbalance in the Nws is also dictated by the polarization of $s$ and $p$ orbitals and their
contributions in addition to that of the contributions of the $d$ orbitals. The variations in the magnetic moments 
(Table-\ref{t-bind}) of the Nw structures investigated arise from the amount of slightly negative polarization of delocalized 
$s$ and/or $p$ orbitals which is listed in Table-\ref{t-spin}. 
Linear Nw has the largest spin imbalance in $s$ and $d$ orbitals amongst the 2D structures we investigated, 
and this accounts for its large net magnetization.
We notice that, all rectangular nanowires show a small negative $p$ polarization.  The unusual magnetic 
behavior of RT-2 is on account of it being the only rectangular Nw with a negative $s$ polarization in 
addition to the negative $p$ polarization.  The contributions from the 
$d$ orbitals decrease until the 3 row ``magic structure" and then increase again as the number of rows 
in the Nws are increased. Hence, the magnetic moments show an initial decrease and then increase again
as the number of rows is increased, in rectangular nanowires. The increased coordination of the `C' type of atoms 
with an increase in the number of rows, reduces their magnetic moments. Increment in the coordination saturates 
slowly as the number of rows is increased, and hence magnetization values converge slowly to their single film limit.

 
The parallelogram Nws with $> 2$ rows of atoms, show negative polarization in the 
$s$ and $p$ orbitals of atoms in the non terminal rows. The amount of the negative polarization 
in the Nws is symmetric about the center of the nanowire and decreases slightly as the number of 
rows is increased. The $d$ contributions do not change as the number of rows is increased and hence
the total magnetic moment of atoms in parallelogram shaped wires decreases with an increase in the
number of rows of atoms.
Although, the negative polarization of $s$ and $p$ orbitals is significantly smaller in value, in comparison with the 
positive polarization of $d$ orbitals, it plays a significant role in the magnetic behavior of the 
parallelogram Nws. 
The increased coordination of `I' and `C' type of atoms leads to a decrease in the charge differences 
in majority and minority spins, thus, reducing their magnetic moments. 
In addition, the negative $s$ and $p$ polarization of the `I' and `C' (Table-\ref{t-spin}) atoms although small, 
reduces the total spin imbalance, resulting in the reduction of net magnetization. 
As a result, the parallelogram Nws show decreased magnetization as the number of rows is increased.
The negative polarization of `I' and `C' atoms in rectangular Nws ($-0.01$ and $0.00$ ) is smaller in comparison with that 
of parallelogram Nws ($-0.02$ and $-0.03$) (Table-\ref{t-spin}). This enhances the net magnetic moment of atoms in rectangular 
nanowires in comparison with parallelogram Nws.

\begin{table*}
\caption{The difference between up and down charge per atom for the  terminal (T), middle (M) and central (C) row atom in parallelogram and rectangular nickel nanowires.}
\begin{tabular} {l c  c  c  c  c  c  c  c  c  c} \hline \hline

Rows & \hspace {0.1in} site \hspace {0.1in} &  \multicolumn{4}{c} { Rectangular Wires } &   &   \multicolumn{4}{c} {Parallelogram Wires }  \\ \cline{3-6}  \cline{8-11}
in Nws  &       & \hspace {0.2in} $s$ \hspace {0.2in} & \hspace {0.2in}  $p$ \hspace {0.2in} & \hspace {0.2in} $d$ \hspace {0.2in}  & \hspace {0.2in} total \hspace {0.2in} & \hspace {0.3cm} & \hspace {0.2in} $s$ \hspace {0.2in} & \hspace {0.2in} $p$ \hspace {0.2in} & \hspace {0.2in} $d$ \hspace {0.2in}  & \hspace {0.2in} total \hspace {0.2in} \\ \hline
\hline

1  & T  & 0.03  & 0.00 & 1.08 & 1.11 &   & 0.03 & 0.00 & 1.08 &1.11  \\
\hline
2  & T  & 0.00  & 0.00 &0.90  & 0.89 &   &0.01  & 0.00 & 0.89 & 0.90 \\
\hline
3  &  T & 0.01  & -0.01 & 0.84 & 0.84 &   &0.00  & -0.01 & 0.87 & 0.86 \\

  &  C  & -0.01  &-0.01  & 0.91 &0.89  &   & -0.01 &-0.03  &0.89  & 0.85 \\
\hline
4  & T  & 0.01  & -0.01 &0.87  &0.86  &   & 0.00 & 0.00 & 0.89 & 0.89 \\

  &  C &  0.00 & -0.01 & 0.85 & 0.84 &   & -0.01 & -0.03 &0.89  &0.86  \\
\hline
5  & T  & 0.01  &-0.01  & 0.89 & 0.89 &   &0.01  &0.00  &0.89  & 0.90 \\

  &  I  &  0.00 & -0.01 & 0.89 & 0.88 &   & -0.01 & -0.02 & 0.89 & 0.86 \\

  &  C &  0.01 & -0.00 & 0.84 & 0.86 &  & -0.01 & -0.02 & 0.84 & 0.80 \\
\hline
6  &  T & 0.01  & -0.01 & 0.90 & 0.90 &   & 0.01 & 0.00 & 0.89 & 0.90 \\

  &  I  & 0.00  & -0.01 & 0.90 & 0.89 &   & -0.01 & -0.02 & 0.89 & 0.86 \\

  &  C  & 0.01  &0.00 & 0.85 &0.86  &  & -0.01 & -0.02 & 0.84 & 0.80 \\
\hline

\end{tabular}
\vskip 0.1in
\label{t-spin}
\end{table*}

In RT-2 Nw, the `C' atom, besides being more coordinated than `T' atom, gives large spin imbalance due to $d$ states. 
This spin imbalance due to $d$ states in RT-2 Nw enhances the net magnetic moment of the atoms belonging 
to the central row. 
This is the only structure, whose terminal atoms show reduced magnetic moments in comparison with 
the magnetic moment of the central atom. 
This reduction is similar to the reduction in the magnetic moments of surface atoms reported  earlier for 
Ni (001) multilayer slabs, with existence of ``magnetically dead layers".\cite{libbermann}
In RT-2 structure, the resultant magnetic moment due to `C' and `T' atoms gives large dip in the magnetization 
curve (plotted in Fig. \ref{f-mag}).
 \begin{figure}
    \centering
    \includegraphics[width=2.2in,angle=270]{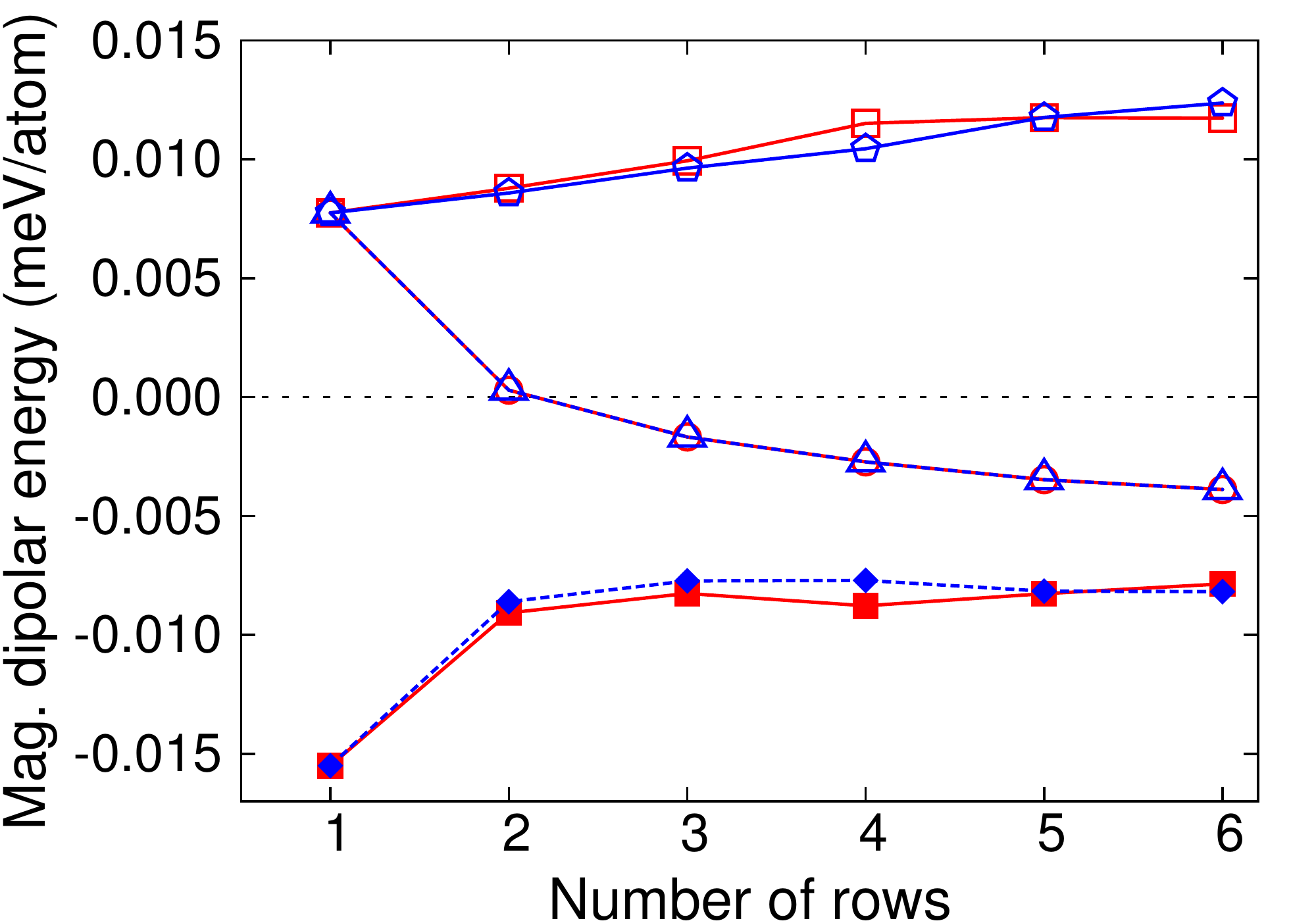}
    \caption{The magnetic dipolar energy for rectangular and parallelogram NWs. anisotropy energy (MAE) for parallelogram and rectangular shaped nanowires as a function of number of rows of atoms. Filled squares, and filled rhombus shows E$_{d}$ for parallelogram and rectangular NWs with magnetization along $\hat{x}$, empty triangles and circles show E$_{d}$ for parallelogram and rectangular NWs with magnetization along $\hat{y}$, empty pentagons and empty squares show E$_{d}$ for parallelogram and rectangular NWs with magnetization along $\hat{z}$.}
    \label{f-di}
\end{figure}
The spin density (not shown here) of the terminal row atoms have significant contribution than that of intermediate and central row atoms, because of spin imbalance due to low coordination. 
 The spin density distribution is symmetric in the $\hat{y}$ direction for parallelogram, rectangular and centered  hexagonal structures. 
 As the number of rows in the parallelogram Nws increase, the negative spin density in the interstitial region gradually increases from the center of the Nw to the terminal row. As discussed in Sec. (\ref{sec-magmom}), this leads to reduced magnetic moment in PR-4 structure in comparison with RT-4 structure, which exhibits negative spin densities only on the terminal row atoms. \\
 
\subsubsection{Magnetic anistropy energy}\label{sec-mae}

Besides a qualitative understanding of the changes in the magnetic moments as the number of rows is  
increased, we are interested in knowing the preferred spin orientations in these nanowires from an
applications perspective. 
 At  reduced dimensions, the asymmetry in the structure has a significant contribution to the shape anisotropy along with the magnetocrystalline anisotropy. The magnetic shape anisotropy was determined from the classical magnetic dipole interaction between the magnetic moments in the system. Figure \ref{f-di} shows the magnetic dipolar interaction energy along the Cartesian axes. One dimensional linear nanowire has the lowest energy when the magnetic moments are aligned along the wire directions. The magnetic moments along the perpendicular direction to the wire plane are unstable, as the dipolar energy is positive.
This is in agreement with earlier work.\cite{tung} The magnetic dipolar energy of the medium axis 
reduces with  increasing number of rows in the nanowires, but is still higher than that of the easy axis of rotation. All nanowires have the hard axis
aligned perpendicular to the wire plane and the magnetic dipolar energy of the hard axis increases as the number of rows in the nanowires are increased.

\begin{figure}
    \centering
    \includegraphics[width=2.2in,angle=270]{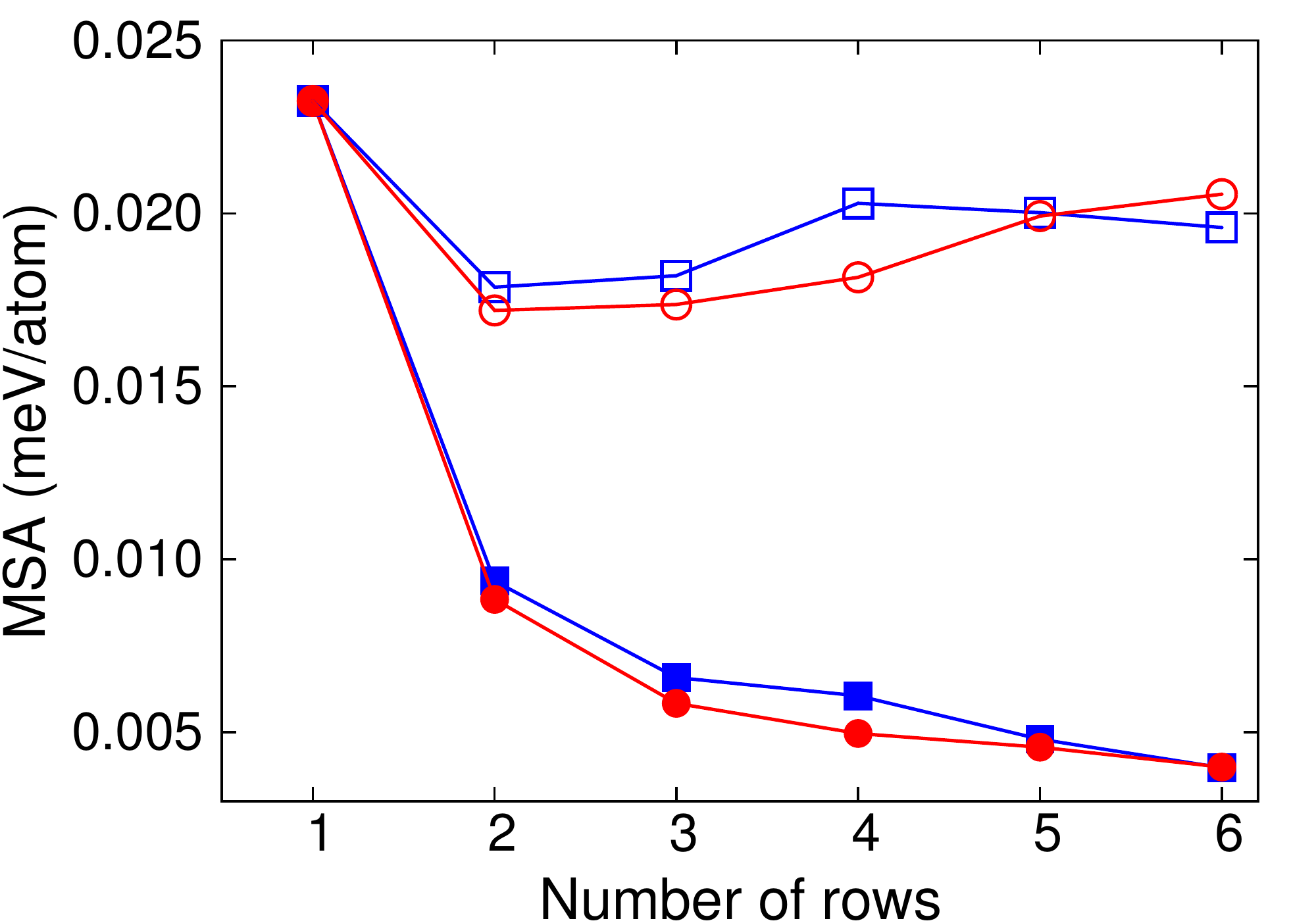}
    \caption{The magnetic shape anisotropy (MSA) energy for rectangular and parallelogram NWs. The squares represents parallelogram NW, whereas, circles show rectangular NWs. Filled legends represent shape anisotropy energy $K_{1}^{s}$, whereas the empty legends show $K_{2}^{s}$. Triangle represents MSA for linear nanowire.}
    \label{f-sa}
\end{figure}

\begin{figure}
    \centering
    \includegraphics[width=2.2in,angle=270]{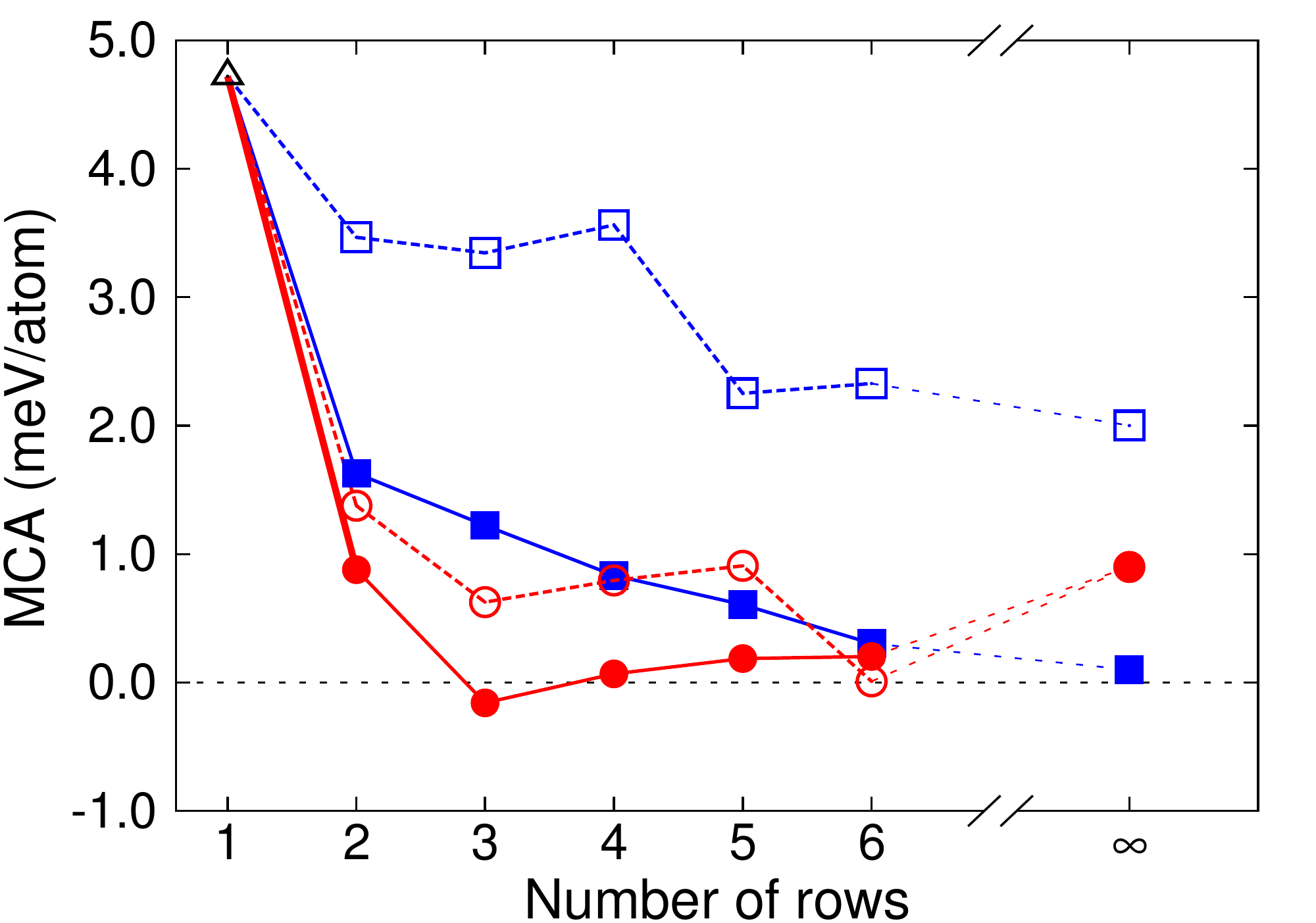}
    \caption{The magnetocrystalline anisotropy energy (MCA) for parallelogram and rectangular shaped nanowires as a function of number of rows of atoms. Squares denote the parallelogram nanowires, whereas circles denote the rectangular nanowires. The filled legends represent $K^{c}_{1}$, whereas empty legends represent $K^{c}_{2}$. Triangle represents MCA of linear nanowire.}
    \label{f-mca}
\end{figure}

The magnetic shape anisotropy $\varepsilon_{d}$ for Ni NW's is shown in Fig. \ref{f-sa}. The shape anisotropy energy $K_{1}^{s}$ decreases sharply 
from linear wire to two row nanowires by ~0.006 meV/atom. Further increase in the number of rows, gradually decreases the anisotropy energy.
The MSA energy $K_{2}^{s}$ decreases from linear nanowire to double row rectangular/parallelogram structure, however, for higher number of rows, the value of MSA is enhanced. This indicates that, as the nanostrip becomes thicker, the magnetic moments are unlikely to orient in the $\hat{z}$ direction due to an increase in the shape anisotropy energy.
The magnetic anisotropy energy (MAE) has another contribution from magnetocrystalline anisotropy (MCA), which originates from the spin orbit coupling in the system. Our calculations show that, the MCA energy is about 200 times larger than the shape anisotropy energy, and hence has a strong impact on the total magnetic anisotropy energy.  MCA for parallelogram and rectangular Nws is plotted in  Fig. \ref{f-mca} . It is interesting to see that $K^{c}_{1}$ for parallelogram structures is found to decrease as the number of rows is increased in the Nw. 
However, for rectangular structures, $K^{c}_{1}$ reduces from RT-1 to RT-2 Nw, but, further 
increase in the number of rows increases the MCA. This behavior of magneto-crystalline anisotropy energy is qualitatively similar to that of the net magnetization (Fig. \ref{f-mag}), as the number of rows is  increased.
$K^{c}_{2}$ for parallelogram and rectangular nanowires has a higher MCA than $K^{c}_{1}$ of respective nanowires, implying it is energetically expensive to have magnetic moments directed in vacuum 
along the $z$ direction.
 As the number of rows (i.e., thickness) of the parallelogram Nws increase, the energy required to rotate the easy magnetization axis is reduced and converges to 0.1 meV/atom for the (111) film.
 For (100) film, due to rotational invariance at 90$^{\circ}$ in the film plane, we observe that the axial magnetization is isotropic in the \textit{XY}-plane ($K^{c}_{1}$=$K^{c}_{2}$=0.9 meV/atom). However, the hard axis remains perpendicular to the plane.
 
\begin{figure}
    \centering
    \includegraphics[width=2.2in,angle=270]{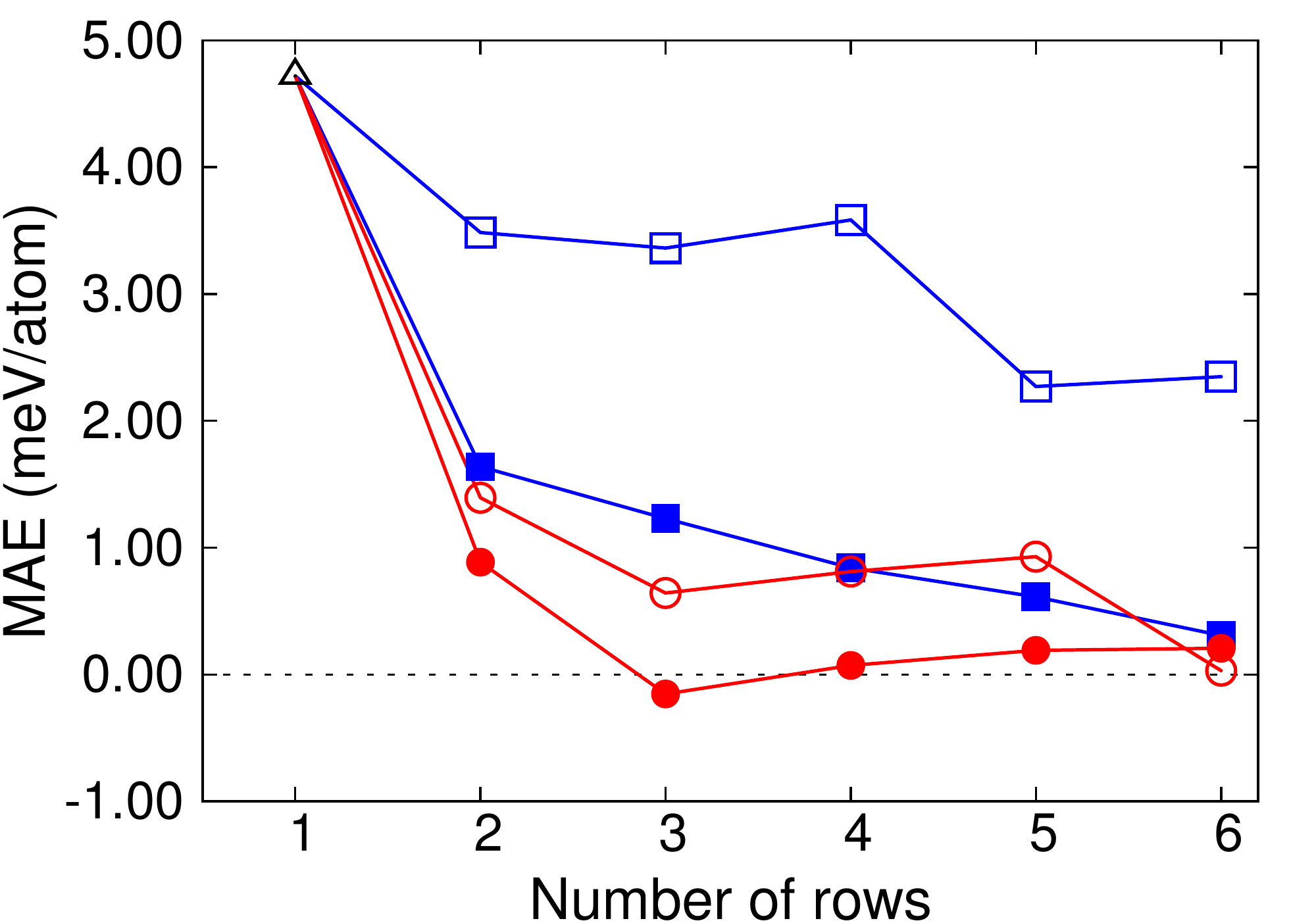}
    \caption{The total magnetic anisotropy energy (MAE) for rectangular and parallelogram NWs. The squares represent parallelogram nanowire, whereas, circles show rectangular nanowires. The filled legends denote the total MAE $K_{1}$, whereas empty legends show $K_{2}$. The MAE of linear nanowire is shown by triangle.}
    \label{f-mae}
\end{figure}

The total magnetic anisotropy energy (MAE) is plotted in Fig. \ref{f-mae}. The MAE ($K_{1}$ and $K_{2}$) is dominated mainly by $K^{c}_{1}$ and $K_{2}^{c}$ values. The linear nanowires prefer to magnetize along the wire axis, in agreement with earlier 
results.\cite{tung, navas} 
However, our result for PR-1 Nw is in contradiction with the reported results\cite{tung}  as we observe that the easy axis of  
magnetization points along the wire direction and not perpendicular to the wire direction. 
Our calculations show that, in rectangular and parallelogram nanowires, the easy axis is along the wire direction, 
except for the magic structure ``RT-2", where, the easy axis is perpendicular to the wire direction and 
in the geometrical $XY$-plane of the structure. We mention that, in case of RT-2 structure, although the magnetic dipolar energy shows 
the preferred magnetization along the wire direction, the energy due to interaction between spin and orbital motion of electrons has a much stronger contribution in the MAE, which leads to the magnetization axis  being perpendicular to the wire axis. 

 In general, we observe that the axial MAE, $K_{1}$ is always larger than $K_{2}$, in all the nanowires. Thus, the external energy required to rotate the magnetization axis perpendicular to the plane is much higher than the energy required
 to rotate in the plane but perpendicular to the wire axis. \\

\begin{table}
\caption{The ballistic anisotropic magnetoresistance ($BAMR$) for linear, parallelogram and rectangular nickel nanowires. ``A" and ``B" show the $BAMR$ (in \%) with respect to \textbf{M} $\parallel \hat{y}$, and  \textbf{M} $\parallel \hat{z}$, respectively, with the current direction being parallel to $\hat{x}$}
\begin{tabular} {l c c c c c c c c c} \hline \hline

Rows  & & \multicolumn{2}{c}{Linear Nw}&  & \multicolumn{2}{c}{Parallelogram Nws} &  & \multicolumn{2}{c} {Rectangular Nws} \\ \cline{3-4}  \cline{6-7} \cline {9-10}
in Nws & & A & B &  &  A & B &  & A & B  \\
\hline

1 &  & 14 & 14 &  & - & - &  & - & - \\
2 &  & - & - & & 0 & 0 & &  0 & 0  \\

3 &  & - & - & & -25 & -25 & & 17 & 0 \\

4 &  & - & - & & 9 & 29 & & 0 &  0  \\

5 &  & - & - & & 0 & 27 & & 19 & 19 \\

6 &  & - & - & & 0 & 13 & & 9 & 9 \\

\hline
\end{tabular}

\vskip 0.1in
\label{t-bamr}
\end{table}

 \begin{figure*}
    \centering
    \includegraphics[width=3.2in,angle=270]{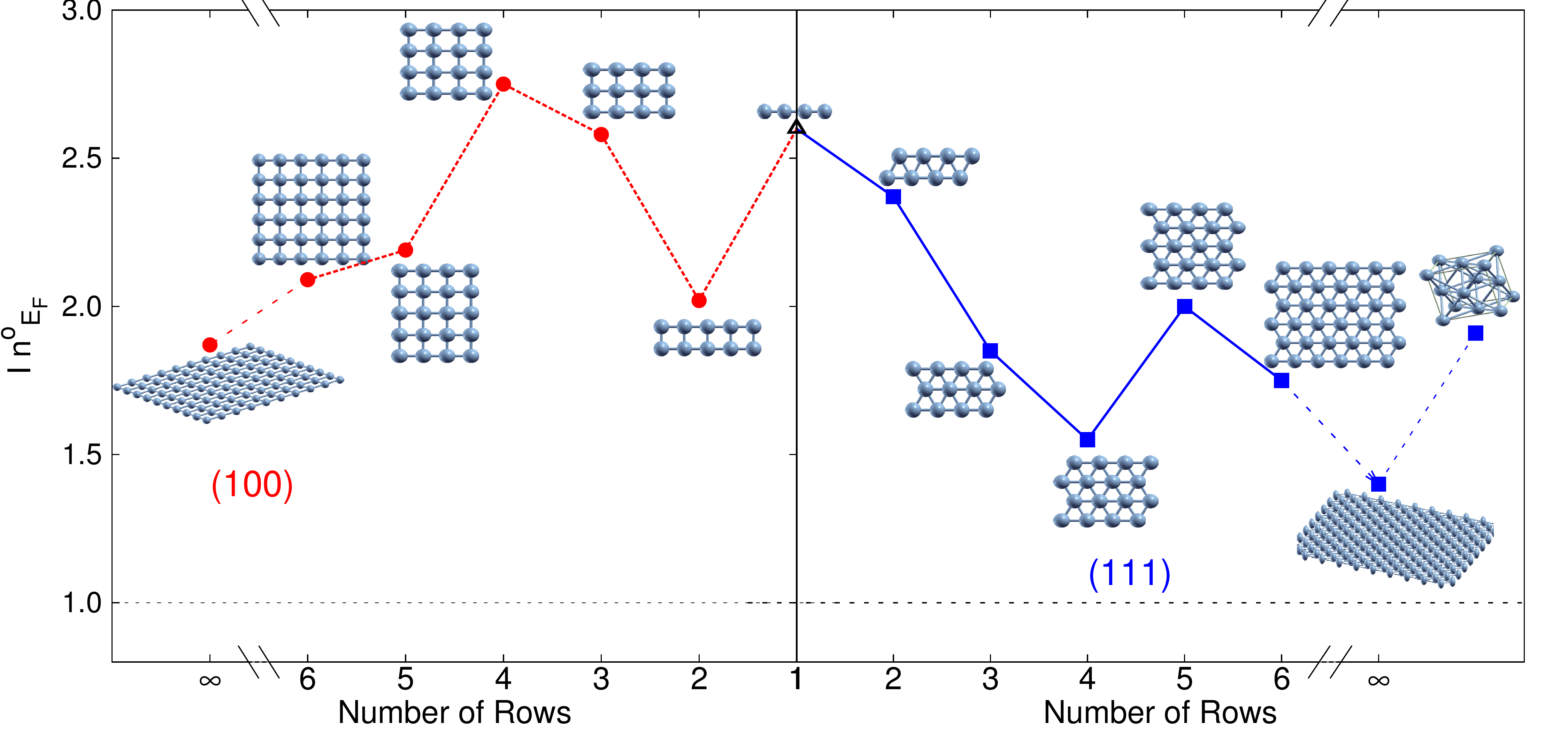}
    \caption{The Stoner criterion for parallelogram and rectangular nanowires as a function of increasing number of rows. Filled squares show parallelogram Nws and filled circles show rectangular Nws. The triangle represents linear nanowire. The dotted horizontal line denotes the Stoner condition $I  n^{o}_{E_{F}}$=1.}
    \label{f-stoner}
\end{figure*}

\subsubsection{Ballistic anisotropic magnetoresistance (BAMR)}

The MAE also has a significant effect on the ballistic transport in the nanostrips. The ballistic conductance is influenced by spin 
orbit coupling and exhibits anisotropy with the magnetization direction.\cite{sokolov} There is a change in the ballistic conductance when the magnetization direction changes with respect to the current direction as the number of bands crossing the Fermi level changes due to the spin
orbit coupling interaction.\cite{velev} We have investigated the size dependence of $BAMR$ in parallelogram and rectangular nanostrips with increasing number of rows. We have calculated the $BAMR$ when the magnetization direction is parallel to $\hat{y}$ and 
parallel to $\hat{z}$ and list these results in Table-\ref{t-bamr}. Our result on $BAMR$ in linear wire is in agreement with the results of Sokolov \textit{et al.}\cite{sokolov} and Velev \textit{et al.}\cite{velev} 
In addition, we find that for the magnetization direction \textbf{M} $\parallel \hat{y}$ and \textbf{M} $\parallel \hat{z}$, the number of conduction channels are 
identical, thus giving the same $BAMR$ ratio in both the cases.  In general, the Ni 
nanostrips exhibit a nonlinear behavior in $BAMR$ with increasing widths.  The nanostrips exhibit finite size fluctuations in 
$BAMR$. The double row structures exhibit no change in conductance with change in magnetization directions. PR-2 is the only structure
with a negative $BAMR$. 
In rectangular NWs, the $BAMR$ with \textbf{M} $\parallel \hat{y}$  and \textbf{M} $\parallel \hat{z}$ are identical except for the RT-2 structure, which has zero $BAMR$ with
\textbf{M} $\parallel \hat{z}$. In rectangular Nws, the magnetization direction does not affect the conductance properties as strongly as in the parallelogram shaped wires, where the $BAMR$ ratio is sensitive to the in plane and out of plane perpendicular orientation of the 
magnetization field.


\subsubsection{Stoner's criterion in Ni Nws}

The magnetism in 3$d$ transition metals such as Ni and Co is explained on the basis of theory of itinerant electrons.\cite{stoner1,slater1} Our spin DFT results for Ni Nws  have  the ferromagnetic ground states to be lower in energy than antiferromagnetic and non magnetic states. 

We have analyzed the appearance of magnetic ground state using Stoner's model.\cite{stoner1,zellar,Marcus,stollhoff}
The product $I\cdot n^{o}_{E_{F}}$ is plotted for Ni Nws, monolayer films (Ni (111) and Ni (100)) in Fig. \ref{f-stoner} and the values of the exchange splitting and normalized DOS are listed in Table-\ref{t-stoner} for comparison. 
The dotted horizontal line in Fig. \ref{f-stoner} shows the product $I\cdot n^{o}_{E_{F}}=1$, which is the demarking value for ferromagnetic behavior. 
The value of the product $I\cdot n^{o}_{E_{F}}$ depends on the Stoner integral which is derived from the exchange splitting in the ferromagnetic (or antiferromagnetic) system. It has been shown that, the Stoner parameter ($I$) does not depend on the atomic arrangement in the system.\cite{zellar}  However it can be seen from Table-\ref{t-stoner}, the Stoner parameter is sensitive to the \textit{dimensions} of the system. In general, the Stoner parameter for rectangular nanowires exhibits larger values than that of the parallelogram nanowires, when compared for the respective number of rows. The value of $I$ for (111) and (100) monolayer film  is small than that of the bulk Ni ($I\cdot n^{o}_{E_{F}}=1.91$). Our calculation  of the Stoner criterion in bulk Ni is in agreement with the previous results.\cite{christensen}


The coupling between the first nearest neighbors in infinite Ni atomic chain ($J_{01} > 0 $) indicates that, the noncollinear magnetic alignment is not stable,\cite{JCTung2} however in small finite zigzag Ni chains, noncollinear magnetic ground states have been observed.\cite{ataca}
Our spin spiral relativistic calculations are in agreement with the results on infinite chains. The strong magnetic anisostropy energy in comparison with the exchange splitting energy enforces the magnetic moments in the infinite Ni nanowires to orient in ferromagnetic order.

iThe least stable linear Nw has the  product $I\cdot n^{o}_{E_{F}}$ larger than all parallelogram nanowires including Ni (111) monolayer and Ni  bulk. Among rectangular Nws, RT-3 Nw exhibits the largest value of the product $I\cdot n^{o}_{E_{F}}$ (2.75).`:w
 The value of $I \cdot n^{o}_{E_{F}}$ is a measure of strength of ferromagnetism in system. 
With an increase in the number of rows in (111) direction, the strength merely decreases with an increase in the number of rows upto 4, however it shows irregular nature for number of rows $> 4$. The monolayer film shows $I\cdot n^{o}_{E_{F}}=1.40$ indicating clear ferromagnetic nature and is in agreement with the previous results.\cite{freeman1,Wu1999498}    
As the number of rows are increased in (100) direction, the $I\cdot n^{o}_{E_{F}}$ shows a dip at RT-1 structure  and a maximum at RT-3 Nw. The product  $I\cdot n^{o}_{E_{F}}$ decreases with  an increase in the number of rows. However,  similar to (111) oriented structures, all rectangular nanowires exhibit ferromagnetic behavior along with Ni (100) monolayer.

Existence of magnetically dead layers of Ni when deposited on non-magnetic substrates viz., Cu, Al and Pd has been described earlier,\cite{libbermann,dtpierce,bergmann,Meservey,Gradmann1977,freeman1,Jarlborg} when the number of atomic layers is less than 3 or 4.
However, our calculations for freestanding monolayers show that, the systems retains ferromagnetism when  not interacting with any substrate. This result points out the role of hybridization of the nonmagnetic substrate atoms in contact with the film/slabs. The previous results show that, in case of pure Ni surfaces, magnetically dead layers are not observed, however, the surface layers show enhanced magnetic moments.\cite{Erskine,CSWang,Eberhardt} In analogy with this, our results for Ni nanowires demonstrate that, the 2D Ni Nws exhibit enhanced magnetic moments than that of bulk, and show an absence of dead layers.


The ferromagnetic strength calculated from Stoner's criterion (Table- \ref{t-stoner}) is larger for Ni Nws in comparison with the bulk Ni. In addition, as discussed in Sec. (\ref{sec-magmom}), the magnetic moments in Ni Nws are enhanced than that of bulk. These two properties can be correlated with the enhanced magnetic anisotropy energy  (\ref{sec-mae}) for Ni Nws in comparison with the bulk.  The large magnetic anisotropy energy in Ni Nws in comparison with the bulk, forces the magnetic moments to couple ferromagnetically and requires high energy to fluctuate /decouple the neighboring magnetic moments from the stable ferromagnetic alignment. This is in agreement with the Mermin-Wagner's theorem, that 2D Ni Nws can show ferromagnetic ground state, because these Nws do not show isotropic Heisenberg interaction, instead they show very large energy differences for different orientations of the magnetic moments.


\begin{table}
\resizebox{3.2in}{!}{
\begin{tabular}{l c c c c}
\hline
System ~~~ &~~~ $\Delta_{ex}$ ~~~& ~~~I ($\Delta e_{x}/M$) ~~~ & ~~ $n^{\circ}_{E_{F}}$ ~~~ &~~~~ $I\cdot n^{\circ}_{E_{F}}$\\
\hline
LR & 0.966   & 0.84 & 3.11 &2.60\\
\hline
PR-1 & 0.680   & 0.72 & 3.27 & 2.37 \\
PR-2 & 0.828   & 0.96 &  1.93 & 1.85\\
PR-3 & 0.673   & 0.78 &  2.00& 1.55 \\
PR-4 & 0.761    & 0.89 &  2.25 & 2.00 \\
PR-5 & 0.704   & 0.82 &  2.14 & 1.75 \\
\hline
RT-1 & 0.920   & 1.02 &  1.98 & 2.02 \\
RT-2 & 0.686   & 1.07 &  2.40 & 2.58 \\
RT-3 & 0.925   & 1.12 &  2.46 & 2.75 \\
RT-4 & 0.901   & 1.05 &  2.09 & 2.19 \\
RT-5 &  0.870  & 0.98 &  2.14 & 2.09 \\
\hline
Ni(111) ML &  0.695  & 0.89 &  1.59 & 1.40 \\
Ni(100) ML & 0.775   & 0.92 &  2.02 & 1.87 \\
\hline
Bulk & 0.635  &  1.01 &  1.91 & 1.91 \\
\hline
\end{tabular}
}
\caption{The exchange splitting ($\Delta_{ex})$ in eV, the Stoner parameter $I$, the normalized value of number of states at $E_{F}$ and the product $I\cdot n^{\circ}_{E_{F}}$ for Ni Nws, monolayer films (Ni (111) and Ni (100)), and bulk, respectively.}
\label{t-stoner}
\end{table}

\section{Conclusion}
We have performed systematic investigations on infinite length, nickel nanowire structures of increasing widths. Our investigations
show that the parallelogram
structure is the most stable Nw geometry. Rectangular Nws have a lower stability in comparison with structures that incorporate 
parallelogram motifs in their geometry. The stability of the parallelogram motif is due to the increased coordination of atoms
and due to contributions from $p$ electrons. In a particular geometrical structure, Nws of higher widths have a higher stability. 
Our studies on the conductance behavior of the Ni Nws show that more stable structures have a lower conductance as compared to the 
less stable structures. All rectangular nanowires have higher spin dependent conductance than the stable parallelogram nanowires. 
In any geometry, the number of spin down conductance channels is more than the spin up conductance channels and the number of 
spin down conduction channels increases substantially with increasing widths of the Nws. 

The parallelogram and rectangular Nws have different magnetic behaviors with increasing widths. Except for the RT-2 structure, the 
magnetic moments of the Nws are enhanced as compared with the bulk and single layered (111) and (100) films.  While magnetization 
decreases with small fluctuations in parallelogram Nws towards their (111) monolayer limit, in rectangular Nws, it initially 
decreases with RT-2 having the smallest magnetic moment and then increases. 
RT-2 exhibits a magnetic anomaly with a smaller magnetic moment in comparison with all other higher coordination rectangular
NWs as well as Ni (100) monolayer. This behavior is contrary to the observation that magnetic moments decrease as the 
coordination of atoms increases.  It is not clear from the small widths investigated here, 
how the magnetic moments in rectangular wires converge to their (100) monolayer value and further investigations are required
to understand this. 

The decrease in magnetization values in parallelogram Nws is on account of the  negative polarization of $s$ and $p$ orbitals,
that decreases the spin imbalance due to  $d$ electrons.  In rectangular Nws, the RT-2 is the only wire that has small negative 
$s$ and $p$ polarization, as compared to all other rectangular wires which exhibit only negative $p$ polarization. The magnetic 
anisotropy energy $K_{1}$ is lower than $K_{2}$ in all the Nws. The MAE decreases with increasing width of parallelogram nanowires, 
and for rectangular nanowires MAE increases with increasing widths. The easy axis of magnetization is 
 found to be along the periodic direction of the wire, except for RT-2 structure, where it is perpendicular. 
 RT-2 Nw exhibits an anomalous behavior in all magnetic aspects as compared with the other rectangular Nws and hence it is a 
 magic structure. The $BAMR$ ratio has a non-linear behavior with increasing widths and is sensitive to the magnetization orientation 
 in parallelogram shaped nanostrips.

All the nickel nanowires investigated exhibit a ferromagnetic behavior based on the Stoner criterion including (111) and (100) monolayer. We see that the exchange splitting depends significantly on the geometrical structure of the nanowire. The rectangular shaped nanowires exhibit enhanced Stoner exchange parameter than those of parallelogram nanowires. The RT-3 structure exhibits the highest Stoner product amongst all Nws investigated. \\

\section{Acknowledgments}
The computational work was carried out using B.A.R.C. mainframe supercomputers (Ameya, Ajeya and Adhya clusters), Yuva-PARAM at C-DAC Pune and the high performance
computing facility at the Centre for Modeling and Simulation at the Savitribai Phule Pune University. The authors thank Rudolf Zeller for fruitful discussions. V. K. acknowledges 
B.A.R.C. for research fellowship during the course of this work, Institut f\"{u}r Festk\"{o}rperforschung (IFF), 
Forschungszentrum-J\"{u}lich and DST-DAAD program for local hospitality and travel support. V. S. would like to acknowledge the 
Dept. of Science and Technology (DST), BCUD-Savitribai Phule Pune University and the DST-DAAD program for funding support. Y. M. acknowledges support from HGF-YiG program VH-NG-513.

\bibliography{reference}
\bibliographystyle{apsrev}

\end{document}